\documentclass[doublecol]{epl2} 
\usepackage{graphicx}
\usepackage{amsmath}
\usepackage{amssymb}
\usepackage{amsfonts}
\usepackage{bm}
\usepackage[colorlinks=true,allcolors=blue]{hyperref}
\usepackage{color}
\usepackage{ulem}

\title{Anomalous glassy dynamics in simple models of dense biological tissue}

\author{Daniel M. Sussman\inst{1} \and M. Paoluzzi\inst{1} \and M. Cristina Marchetti\inst{1} \and M. Lisa Manning\inst{1}}
\shortauthor{D. M Sussman \etal}

\institute{                    
  \inst{1} Department of Physics and Soft and Living Matter Program, Syracuse University, Syracuse, New York 13244, USA}
  

\abstract{In order to understand the mechanisms for glassy dynamics in biological tissues and shed light on those in non-biological materials,  we study the low-temperature disordered phase of 2D vertex-like models. Recently it has been noted that vertex models have quite unusual behavior in the zero-temperature limit, with rigidity transitions that are controlled by residual stresses and therefore exhibit very different scaling and phenomenology compared to particulate systems. Here we investigate the finite-temperature phase of two-dimensional Voronoi  and Vertex models, and show that they have highly unusual, sub-Arrhenius scaling of dynamics with temperature. We connect the anomalous glassy dynamics to features of the potential energy landscape associated with zero-temperature inherent states.}

\begin{document}

\maketitle

\section{Introduction} 
How do materials become rigid? In materials that crystallize, rigidity coincides with increasing structural order and a corresponding broken symmetry.  A large class of molecular fluids can, however, be supercooled past this crystallization point, and along this path their viscosity $\eta$ increases sharply with no obvious corresponding changes in the structural order. In practice, a material is deemed a rigid glass when either $\eta$ or $\tau_\alpha$, the time scale characterizing the relaxation of density fluctuations, exceeds some threshold value \cite{angell1995formation}. 

In simulations or experiments, the onset of rigidity is typically identified by plotting relaxation time data in an ``Angell plot,'' $\log(\tau_\alpha)$ vs. $T_G/T$ \cite{angell1995formation}, where the glass transition temperature $T_G$ is the temperature at which $\tau_\alpha$ grows past some threshold. Strong glasses appear as straight lines on such plots, and fragile glasses strongly curve upwards. In a simple Arrhenius model for activated dynamics, the slope of the Angell plot is the energy barrier $\Delta E$ required for activation, suggesting that fragile glasses are ``super-Arrhenius,'' with energy barriers that grow as the temperature decreases. This has driven intense experimental and theoretical studies of glassy systems, with a focus on understanding how the growth of $\Delta E$ as the temperature decreases may or may not be a signature of a diverging length scale in the problem \cite{Berthier2011,charbonneau2017glass,Cates2017}. Despite these efforts, the fundamental mechanism behind disordered rigidity transitions, even in very simple systems, is still under active study. 
 
Recent observations of glass-like transitions in dense biological tissues~\cite{angelini2011glass, schoetz2013glassy, nnetu2013slow,oswald2017jamming} may seem even more difficult to explain, given that cells are themselves complex entities. Over the past 20 years scientists have coalesced around simple classes of \textit{vertex} and \textit{cellular Potts} models of dense tissues that have been remarkably predictive of cellular structures and mechanics~\cite{graner1992simulation,Brodland2004,Honda2001,Manning2010,Bi2014}. Recently, these  models have been shown to exhibit features of glass transitions~\cite{kabla2012collective, Bi2016, chiang2016glass}.
A natural question, then, is whether glass transitions in models for biological tissues are similar to or different from those in simple fluids. This question, which is of clear interest to biophysicists, is also pressing from a fundamental theoretical perspective, as vertex models have several unusual features that may help to shed light on the basic mechanisms for glassy rigidity in non-biological systems.

In particular, much work in statistical physics over the past 25 years has focused on understanding how the $T=0$ rigidity transition in particulate systems (called the jamming transition) might inform our understanding of finite-temperature glassy behavior~\cite{LiuRev}. There is significant evidence that the excitations responsible for viscosity in supercooled liquids are strongly correlated with the vibrational normal modes of the $T=0$ inherent states~\cite{widmer2008irreversible, tanguy2010vibrational}. Some vertex-like models also have a $T=0$ rigidity transition~\cite{Bi2015, Sussman2017} but the mechanisms that drive rigidity in these systems are very different from those in particulate matter \cite{Merkel2017, Sussman2017, moshe2017geometric}. In vertex-like models, rigidity transitions can be driven by residual stresses~\cite{moshe2017geometric}, resulting in scaling laws that are different from those in particulate systems~\cite{Merkel2018}. Since the $T=0$ behavior of vertex-like models is interesting and unexpected, it is natural to wonder whether their glassy dynamics is similarly unusual. If so, this may allow us to test the validity of standard assumptions from glass physics that relate inherent states to finite-temperature dynamics.

In this manuscript we focus on 2D vertex-like models that represent a dense tissue as a tiling of space, in which each tile corresponds to a coarse-grained representation of a cell. In vertex models the cells are polygons (or more generally, d-dimensional polyhedra), and the vertices of the polygons are the degrees of freedom. Recent work has identified an alternative for modeling tissue, taking each cell to be a local Voronoi volume \cite{Bi2016,SAMOS}. The degrees of freedom, thus, are the \textit{Voronoi cell centers}, with some evidence suggesting that cell nuclei position themselves at the centers of Voronoi cells defined by their surrounding neighbors \cite{kaliman2016limits,Idema2017}. In both of these vertex-like models, the degrees of freedom are controlled by an energy functional which specifies quadratic constraints tethering each cell to some preferred generalized volume and surface area (e.g. cross-sectional area and perimeter in 2D).

Unlike particulate jamming, which is marginally stable precisely at the transition density \cite{LubenskyRev,LiuRev}, the 2D Voronoi model at $T=0$ is \textit{always} at the marginal point, with precisely balanced constraints and degrees of freedom \cite{Sussman2017}.  The 2D vertex model, on the other hand, has a bona-fide  $T=0$ rigidity transition \cite{Bi2015}. Although the $T=0$ behavior of these two models seems quite different, the 2D Voronoi model has a nearby avoided critical point that may be in the same universality class as that of the vertex model (since a continuous parameter could be used to tune the Voronoi constraints into the underconstrained vertex model). Therefore, it is interesting to compare their low-temperature dynamics to see if they are both dominated by that critical point.

Our goal is to study these models' glassy behavior and compare it to the observations made in simple molecular fluids. Strikingly, we show that these models have very unusual fluid-phase dynamics, with no crossover to standard Arrhenius or super-Arrhenius glassy dynamics even when $\tau_\alpha \gg 1$. In the 2D Voronoi model this can be associated with a low-energy vibrational mode spectrum very different from that seen in particulate matter, which we relate to the marginal $T=0$ phase of the model.

\section{Models and methods}
\subsection{Voronoi model}
We perform dynamical simulations where the forces on each degree of freedom are computed in terms of the Voronoi tessellation derived from the instantaneous cell positions. We note that recently a ``Voronoi liquid'' has been proposed where a force is defined that moves each Voronoi position towards the centroid of the Voronoi cell to which it belongs \cite{ruscher2016voronoi,farago2017anomalous}. Here we study the biologically motivated forces described by gradients of the vertex energy functional described below; it would be interesting to understand any relationship between these two different models.  A dimensionless version of the standard vertex energy functional is \cite{Farhadifar2007,Bi2015,Bi2016}
\begin{equation}\label{eq:energy}
e=\sum_{i=1}^N k_A \left(a_i - a_{0,i} \right)^2+\left(p_i - p_{0,i} \right)^2.
\end{equation}
Here $N$ is the total number of cells, $a_i$ and $p_i$ are the area and perimeter of cell $i$, where we have chosen the unit of length so that the average area of the cells in the simulation $\langle a_i\rangle = 1$. The ``preferred'' or target values of these geometric quantities are $a_{0,i}$ and $p_{0,i}$, and $k_A$ governs the relative area to perimeter stiffness of the cells.

While molecular dynamics codes and meshing codes to compute tessellations of a point set are each individually well-established, until very recently there were only very rudimentary available simulation packages available to combine the two. We have recently developed a hybrid CPU/GPU code, \textit{cellGPU}, which combines the two in a high-performance environment \cite{Sussman2017cellGPU,CGAL}. We use this software to simulate overdamped Brownian dynamics, where the positions of the cells are updated at each time step according to $\Delta r_{i\alpha} = \mu f_{i\alpha} \Delta t + \eta_{i\alpha}$. Here $\Delta t = 0.01\tau$ is the integration time step simulated, where $\tau$ is the simulation time unit given by the inverse friction coefficient $\mu$, $f_{i\alpha}$ is the force acting on cell $i$ in Cartesian direction $\alpha$, and $\eta_{i\alpha}$ is a normally distributed random force with zero mean and $\langle \eta_{i\alpha}(t) \eta_{i\beta}(t') \rangle = 2 \mu T \Delta t \delta_{ij}\delta_{\alpha\beta}$ \cite{branka1999algorithms}. The temperature $T$ sets the scale of translational noise applied to the cells at each time step.

Initial works studying the Voronoi model have focused on monodisperse systems, but just as in particulate settings there are model parameter regimes in which monodisperse Voronoi packings have a tendency to crystallize into hexagonal ground states \cite{Sussman2017,Ciamarra2017}. To maintain our focus of the disordered solid regime, we work with bidisperse mixtures of cells with different preferred geometric quantities. Inspired by disk packings that frustrate the hexagonal lattice \cite{Koeze2016}, we choose a $50:50$ mixture of $N=1024$ cells with $a_{0,\alpha}/a_{0,\beta}=4/3$ for cell types $\alpha$ and $\beta$. We then scale the preferred perimeters of the cells so that the dimensionless target shape parameter $q_0 \equiv p_{0,i}/\sqrt{a_{0,i}}$ is identical for all cells.

\subsection{Vertex model}
In addition to the Voronoi model, we also simulate a thermal version of the vertex model controlled by the same energy functional, applying thermal noise directly to the vertices of each cell \cite{Sussman2017Interface}. Unlike the Voronoi model, which takes care of cell-neighbor exchanges by continuously updating the Voronoi tessellation, in vertex models one must explicitly handle rules for how cells undergo ``T1,'' or neighbor-exchange, transitions. Here we implement a scheme in which T1 transitions occur whenever two vertices are closer than some threshold value, which we take to be $l_c=0.04$. 

For every Voronoi or vertex model state point $(q_0,T)$ of interest we run 100 independent simulations. All simulations are thermalized at their target temperature for $10^4\tau$ before recording data, and we perform simulations of maximum length $10^5 \tau$. Below we restrict our analysis to systems with a characteristic relaxation time $\tau_\alpha \lesssim 10^4$. Given the qualitatively similar behavior we find between the vertex and Voronoi models, below we focus on presenting data for the Voronoi model, and show our vertex model results in the supplemental material.

\section{Dynamical properties of the glassy phase}
\begin{figure}
\includegraphics[width=1.0\linewidth]{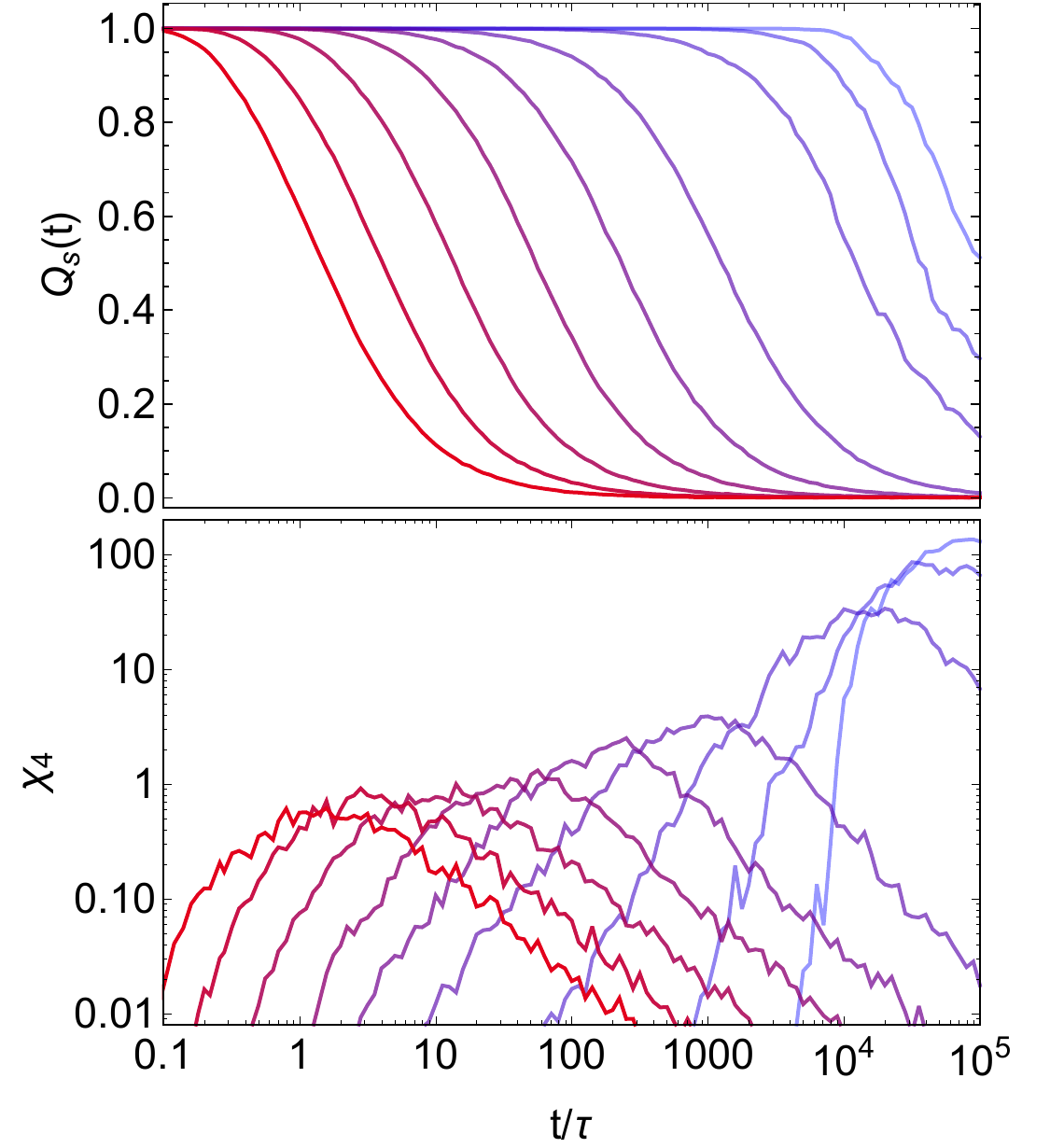}
\caption{\label{fig:selfDynamics} (top) The self-overlap function quantifies the fraction of cells that have moved more than a threshold value. Here we plot $Q_s(t)$ for the bidisperse Voronoi model with $q_0=3.75$, and $T=5.5\times 10^{-4} - 1.5\times 10^{-1}$ (light blue to dark red). (bottom) The four-point susceptibility $\chi_4$ is measured for the same parameters, with a slowly growing peak value.}
\end{figure}

We begin by characterizing the relaxation time as a function of temperatures and preferred shape using the decay of the self-overlap function \cite{keys2007measurement}. This function measures the fraction of particles that have been displaced by more than a characteristic distance $b$ after a time $t$,
\begin{equation}
Q_s(t) = \frac{1}{N}\sum_{i=1}^N w \left( |\vect{r}_i(t) - \vect{r}_i(0) | \right),
\end{equation}
where $\vect{r}_i$ is the vector position of cell $i$, $w$ is a window function, $w(r \leq b) = 1$ and $w(r>b) = 0$. The cutoff $b$ plays a very similar role to a choice of wavevector when looking at the decay of the self-intermediate scattering function,
\begin{equation}
F_s(q,t)=N^{-1} \left\langle \sum_i e^{ i \vect{q}\cdot \left(\vect{r}_i(t) - \vect{r}_i(0)\right) } \right\rangle.
\end{equation}
Below we follow convention and look only at the decay of the larger cell species, and choose $b$ to be half of a characteristic cell size ($b = \sqrt{a_\alpha}/2$). Figure \ref{fig:selfDynamics} illustrates the behavior of this function for the bidisperse Voronoi model at moderately low preferred shape parameter, $q_0=3.75$. It is clear that the onset of rearrangements and relaxation is progressively delayed as the temperature decreases, just as in simple fluids.

One can also define characteristic scale by looking at the four-point susceptibility, $\chi_4$, shown in Fig. \ref{fig:selfDynamics}. Letting $\delta F_s(\vect{q},t)$ be the difference between the instantaneous value of the scattering function and the mean, one has $\chi_4= N\langle \delta F_s(\vect{q},t) \delta F_s(\vect{q},t)\rangle$, where $\chi_4$ is typically evaluated at a wavevector correspond to the maximum of the static structure factor \cite{szamel2006four}. The growth of the peak of $\chi_4$ as the temperature is lowered, often interpreted as a measure of the size of collectively rearranging regions, is again reminiscent of a standard glassy system and suggestive of a slowly growing length scale that goes from sub-cell scale to perhaps $\sim 2\sqrt{a_\alpha}$.

\begin{figure}
\centerline{\includegraphics[width=1.0\linewidth]{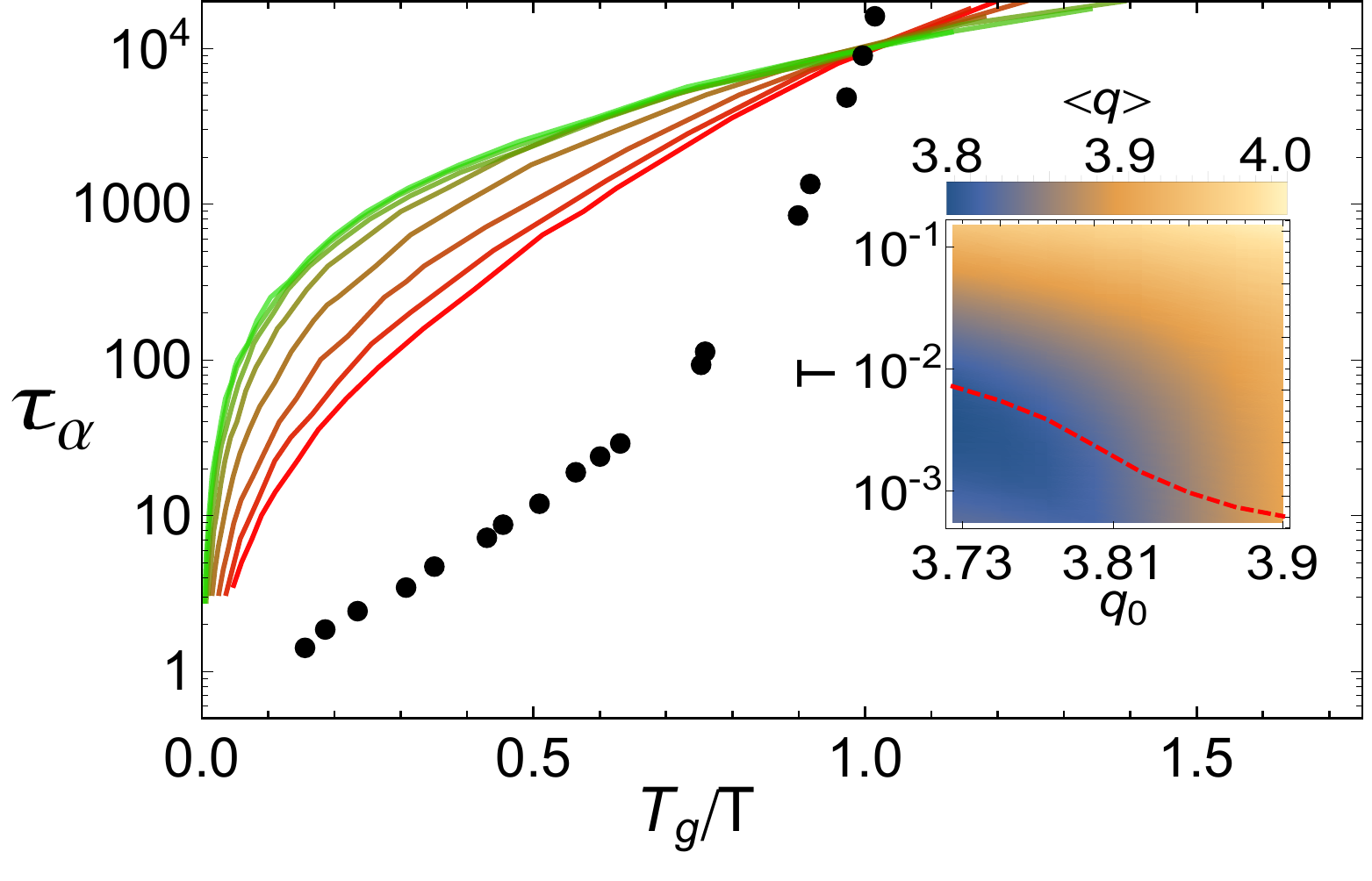}}
\caption{\label{fig:angellPlot} Angell plot of log relaxation time versus inverse temperature in the Voronoi model, normalized so that $T_g$ is the temperature at which $\tau_\alpha=10^4$. The different curves correspond to $q_0=3.725, 3.75,\ldots,3.9$ (dark red to light green). Strikingly, in this representation the Voronoi model exhibits \textit{sub}-Arrhenius scaling of the relaxation time with temperature. We highlight this by also presenting data from a 3D Kob-Andersen mixture of Lennard Jones particles at density $\rho =1.2$ (data from Ref. \cite{berthier2011role}). [Inset] Color map of the mean measured shape parameter, $\langle q\rangle$, as a function of $q_0$ and $T$. The dashed red line corresponds to a line of constant $\tau_\alpha=10^4$.}
\end{figure}

From the decay of the overlap function we define the characteristic relaxation time of the system, $\tau_\alpha$, via $Q_s(\tau_\alpha) = 0.2$, i.e., the time at which most of the large cells have displaced by a magnitude of order their own size. In addition, one can define a structural order parameter, $\langle q \rangle = \langle p_i/\sqrt{a_i}\rangle$. In the vertex model it was shown that this order parameter is an excellent indicator of the $T=0$ rigidity transition \cite{park2015unjamming}. The inset to Fig. \ref{fig:angellPlot} combines these dynamical and structural data, where the mean shape of the cells is encoded in a heat map and a line corresponds to a dynamical transition at which $\tau_\alpha = 10^4$. On this logarithmic scale it is clear that there can be a significant decoupling between the dynamical transition and a structural transition defined purely by measuring static structural properties of the system. In this sense the Voronoi model fits neatly into the standard paradigm of glass-forming liquids, and this decoupling is in keeping with the expectation based on the work in the athermal limit of the monodisperse model \cite{Sussman2017}.

In Fig. \ref{fig:angellPlot} we plot the $\alpha$-relaxation time data in the Angell plot representation discussed in the introduction, and it reveals strikingly anomalous behavior. Remarkably, the Voronoi model displays \textit{sub-Arrhenius} scaling of the relaxation time as the temperature is lowered. This sub-Arrhenius behavior suggests that either typical glassy energy barriers in the Voronoi model \textit{decrease} in colder glasses, or that a local picture of activated dynamics does not hold; both scenarios are highly unusual in the context of standard glassy materials. The results for the vertex model are qualitatively identical, as we show in the supplemental material. This confirms that sub-Arrhenius behavior is a generic feature of these vertex-like models.

\section{Structural properties of the glassy Voronoi phase}
To better interpret the anomalous glassy dynamics reported above, and to understand the connection between these dynamical results and the $T=0$ behavior of the Voronoi model \cite{Sussman2017}, we investigate the vibrational mode structure of the inherent states of our various thermal configurations. In particular, for equilibrated  $(q_0,T)$ state point configurations we use a FIRE energy minimization algorithm to find a nearby energy minima \cite{bitzek2006structural}. We then compute the dynamical matrix, $\mathbf{D}_{ij} = \partial^2 e/\partial \bm{r}_i \partial \vect{r}_j$, corresponding to these energy-minimized configurations. The eigenvectors of this matrix are the vibrational modes with frequencies $\omega_i = \sqrt{\lambda_i}$ where $\lambda_i$ are the eigenvalues.

\begin{figure}
\centerline{\includegraphics[width=1.0\linewidth]{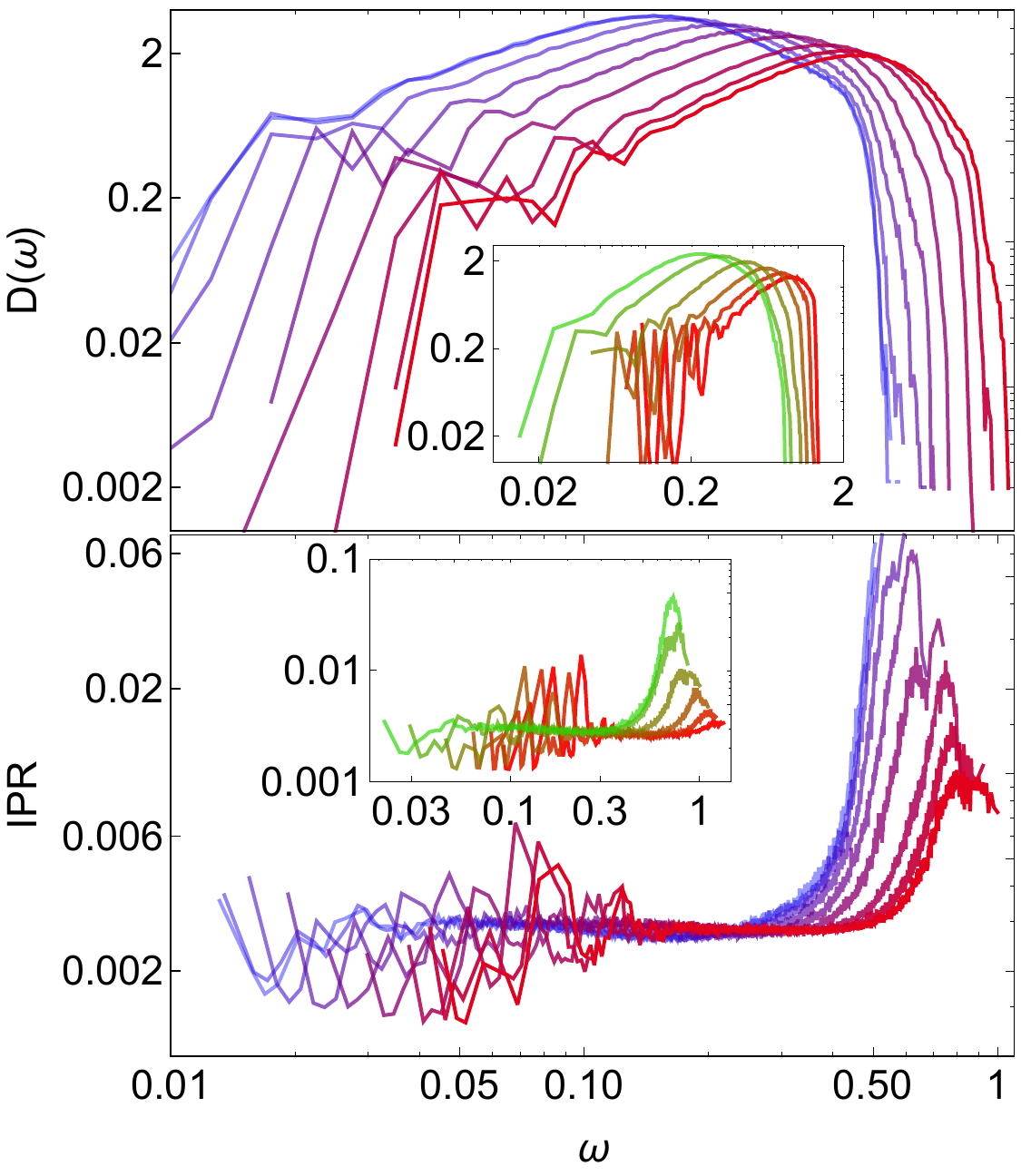}}
\caption{\label{fig:DosIPR} (top) The density of vibrational modes in the 2D Voronoi model. The data in the main frame corresponds to systems with $q_0 =3.8$ and $T=1.05\times 10^{-3} - 1.5\times 10^{-1}$ (light blue to dark red). The data in the inset corresponds to systems with $T=0.3$ and $q_0=3.725, 3.75,\ldots,3.85$ (dark red to light green). (bottom) The average inverse participation ratio for modes of approximately the same vibrational frequency. All colors correspond to the same state points as in the top plot. }
\end{figure}

For all of our data for the Voronoi model we find that the only zero energy modes of the dynamical matrix correspond to the two (trivial) translational modes, so that the energy minima are always mechanically stable. This is in contrast with the vertex model, where the energy landscape becomes flat in many directions above $p_c\approx 3.81$, corresponding to a large number of non-trivial zero modes \cite{Bi2014,Bi2015}. The inset of Fig. \ref{fig:DosIPR} plots the result of quenching from a relative high temperature and varying $q_0$; consistent with the infinite-temperature quenches studied in Ref. \cite{Sussman2017} we find that as $q_0$ increases there are an increasing population of low-frequency modes. 

The main frame of Fig. \ref{fig:DosIPR} reveals a structural feature of these energy minima which is also strikingly unusual. The energy minima of \textit{colder} Voronoi fluids are \textit{softer}, with more low-frequency vibrational modes. This is in stark contrast to the standard picture of the glassy energy landscape, in which deeper energy minima also correspond to higher-curvature basins. 

In models for molecular fluids and particulate matter, researchers have begun to establish a connection between the landscape curvature and energy barriers: lower eigenvalue modes typically have lower associated energy barriers \cite{xu2010anharmonic,karmakar2010statistical}. Although such a connection has not yet been established in vertex-like models, it is natural to conjecture that the flatter landscape we find at low temperatures may give rise to lower energy barriers and sub-Arrhenius dynamics.

However, we find find that the spatial structure of these modes is very different from those in particulate glasses. Specifically, the localized (or quasi-localized) low-frequency modes that are common in particulate glasses are not found in the Voronoi model. In Fig. \ref{fig:DosIPR} we also plot the inverse participation ratio (IPR). The IPR, $Y(\omega)$,  for eigenmode $\vect{u}_i$ is defined by
\begin{equation}
Y(\omega) = \frac{\sum_i^N \left| \vect{u}_i(\omega)\right|^4 }{\left[\sum_i^N  \left| \vect{u}_i(\omega)\right|^2\right]^2 }.
\end{equation} 
This measures the degree of localization for each eigenmode, where $Y=1$ corresponds to a mode localized to a single cell and $Y\sim 1/N$ corresponds to a completely extended mode involving every cell in the system equally. In both jammed \cite{xu2010anharmonic} and glassy \cite{lerner2016statistics,Mizuno2017} particulate systems there is clear evidence that the softest (lowest-frequency) non-plane-wave modes are quasi-localized, involving a core of potentially rearranging particles with an elastic tail. In strong contrast, we see that essentially all of the low-frequency modes in the Voronoi model are collective, much like the disordered modes in the plateau of the density of states in jammed systems. Thus, the density of states supports an interpretation in which the anomalous dynamics result from potentially decreased energy barriers at lower temperatures, and the eigenmode structure suggests that, at least in linear response, localized excitations are not dominating glassy dynamics in these systems.

\section{Discussion}
Through extensive Brownian dynamics simulations we have demonstrated that the low-temperature fluid phase of vertex-like models display very unusual glassy behavior. Although the characteristic relaxation time in the disordered phase becomes very long as the temperature is decreased, this growth is slower than exponential in the inverse temperature. We further studied the inherent states of the 2D Voronoi model after thermalization at many different temperatures; this data was consistent with the unusual dynamics observed and revealed that \textit{colder} glasses were in characteristically \textit{softer} energy minima. 

We show in the supplemental material that replotting Fig. \ref{fig:angellPlot} on a log-log scale suggests that the relaxation time of the Voronoi model has a power-law form even for $\tau_\alpha \gg 1$. The larger-$q_0$ state points show a scaling of $\tau_\alpha \sim T^{-3/2}$ over the entire range of temperatures studied, whereas simulations at lower $q_0$ are perhaps crossing over from power-law scaling at high temperatures to another form at very low temperatures. As seen in Fig. \ref{fig:angellPlot}, though, we never observe a clean Arrhenius scaling over any substantial range of $T$.

Such a power law scaling has been observed in geometrically frustrated versions of the two-dimensional XY model \cite{esterlis2017avoided} and attributed in part to the phenomenon of avoided criticality. It may be fruitful to think of the Voronoi model behavior in terms of an avoided critical point \cite{Sussman2017}. Even though at zero temperature the Voronoi model does not have an unjamming transition it is close to models that do, either by taking $k_A\rightarrow 0$, by moving from flat space to a space of non-zero Gaussian curvature so that disordered Voronoi geometries can satisfy all of their energetic constraints, or by removing the constraints that every cell adopt a Voronoi geometry (i.e., by studying a vertex model). Indeed, in the supplemental we also provide data for the dynamics of vertex model evolving according to the same energy functional, and show that the low-temperature dynamics are similar. 

Although for large $q_0$ there appears to be power-law rather than activated dynamics, we have been, as yet, unable to find an accompanying diverging length scale in the model. At zero temperature we have observed that the mechanical rigidity transition is completely collective, with no signature of growing regions of mechanically stable cells. An important question, then, is whether such a static length scale can be observed and used to formulate a better theoretical understanding of these cellular models -- certainly the apparent absence of activated dynamics suggests that they may be more easily understood than typical glassy systems. This is particularly interesting in light of the slow growth of the peak of $\chi_4$ seen in Fig. \ref{fig:selfDynamics}, which suggests the possibility that nonlinearities are a key to understanding structural length scales in this model.

The statistics of the linear properties of the energy landscape the Voronoi model explores at different temperatures, i.e., the vibrational mode frequencies and their spatial structure, is quite striking. The fact that lower-frequency modes are available to the system at colder temperatures suggests that residual stresses -- which are the mechanism of zero-temperature rigidification in the vertex model and underconstrained versions of the Voronoi model \cite{Merkel2017}, as well as in crystalline Voronoi models \cite{moshe2017geometric} -- may also be crucial to understand the finite-temperature behavior of these models. We speculate that these unusual dynamics may be a signature of broad classes of residual-stress-rigidified systems, such as underconstrained Mikado models with strain-controlled rigidity transitions \cite{vermeulen2017geometry}.

Additionally, the apparently collective behavior of even the low-frequency modes suggests that glassy cell rearrangements may be very different from particulate ones. In the particulate setting quasi-localized vibrational modes have frequencies which are known to strongly correlate with the scale of the energy barriers to motion in the direction of those modes \cite{xu2010anharmonic}. This allows the low-frequency modes to be used as strong predictors of sites of potential plasticity under strain or rearrangements under thermalization \cite{manning2011vibrational}. The fact that we do not observe such quasi-localized modes suggests that either the cellular rearrangements are fundamentally less local in the Voronoi model, or that the tight connection between linear and non-linear behavior seen in particulate settings may not hold here. An important open question involves resolving precisely this issue, either by directly studying the statistics of the energy barriers in the model or by using machine-learning methods to infer such features \cite{cubuk2015identifying,schoenholz2016structural}.

\acknowledgments
DMS and MP contributed equally to this work. We would like to thank Matthias Merkel for fruitful conversations. This work was primarily supported by NSF-POLS-1607416 and the Simons Foundation Targeted Grant 342354 (MCM, MP) and an Investigator grant 446222 (MLM) in the Mathematical Modeling of Living Systems. Additional support was provided the National Science Foundation awards DMR-1352184 (MLM)  and DMR-1609208 (MCM), and a Cottrell Scholar award from the Research Corporation for Science Advancement (MLM). We acknowledge computing support through NSF ACI-1541396 and an XSEDE allocation on \textit{Comet} through Grant No. NSF-TG-PHY170034.

\bibliography{glassyVoronoi_bib}
\newpage

\newpage

\section{Supplemental Material} 
In this supplemental material we provide additional standard characterizations of both the structure and dynamics of the low-temperature Voronoi and vertex models, and briefly comment on our choice of bidisperse mixtures in contrast to the monodisperse models previously studied.

\section{Dynamics}

\begin{figure}
\centerline{\includegraphics[width=0.85\linewidth]{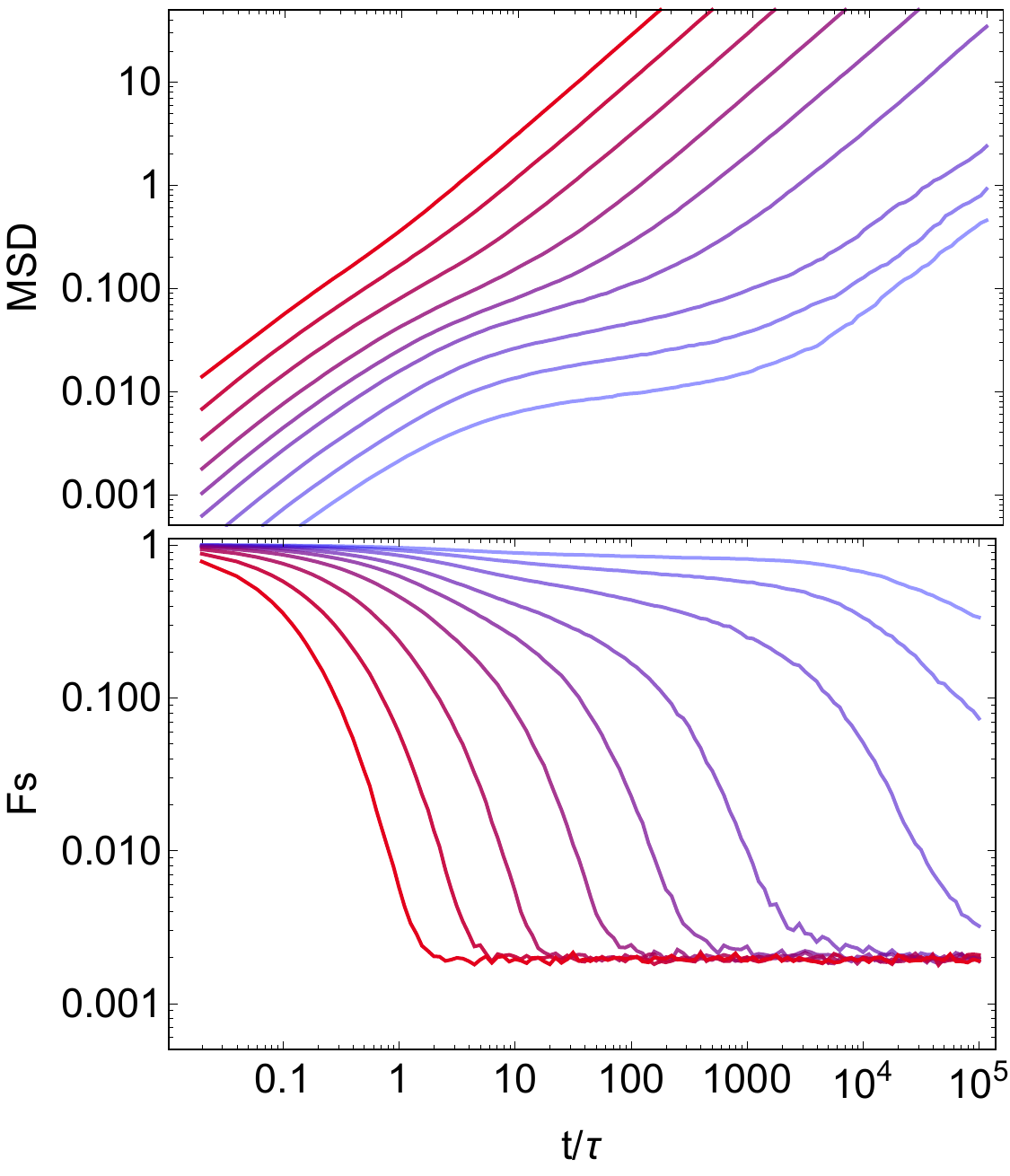}}
\caption{\label{fig:StackedMsdFs} (top) Mean-squared displacement for $q_0=3.75$ and $T=1.05\times 10^{-3} - 1.5\times 10^{-1}$  (light blue to dark red). (bottom) Examples of the self-intermediate scattering function for the bidisperse Voronoi model for the same set of parameters.}
\end{figure}

In the main text we showed the decay of the overlap function and used it to measure the characteristic relaxation time. Here we also directly show the decay of the self-intermediate scattering function in Fig. \ref{fig:StackedMsdFs}; as is typical, the long time decay of the scattering function and of the overlap function give equivalent information, with the scattering function also resolving information about shorter-time decorrelation processes. We also show data for the mean-squared displacement (MSD) of particles in Fig. \ref{fig:StackedMsdFs}. This emphasizes the lack of a true plateau in the MSD,  even when the relaxation time is very long, indicating the absence of strong particle caging. The possible connection between the nearly power-law scaling of the MSD in the quasi-plateau regime and the sub-Arrhenius scaling of the structural relaxation time may be an interesting avenue of future research.

\begin{figure}
\includegraphics[width=0.85\linewidth]{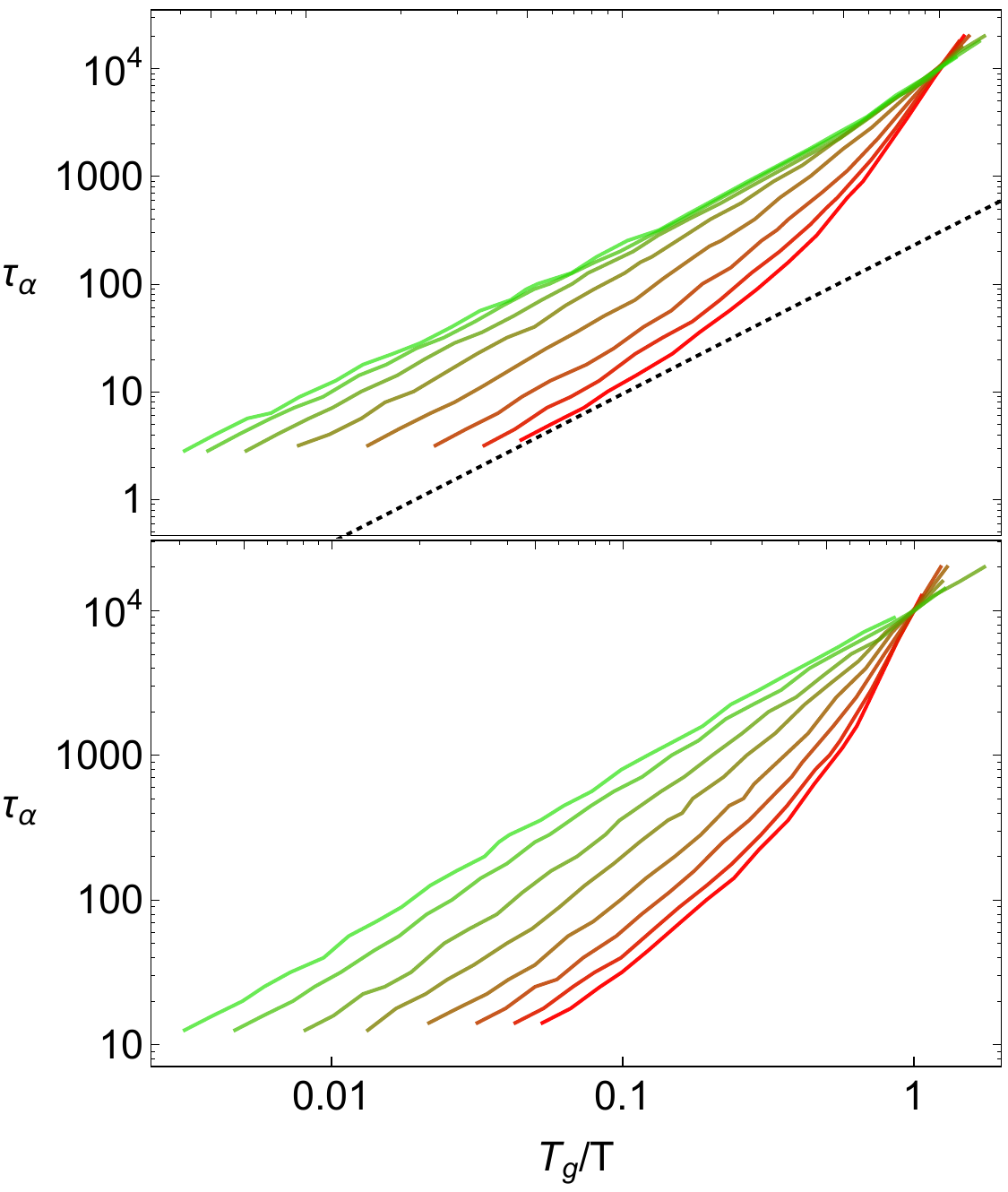}
\caption{\label{fig:angellLogLog} (top) The Voronoi and (bottom) vertex models exhibit approximately power-law scaling of the relaxation time at high temperatures. where the different curves correspond to $q_0=3.725, 3.75,\ldots,3.9$ (dark red to light green). The dashed line is a guide to the eye with slope 1.5.}
\end{figure}

In the main text we presented relaxation time data versus inverse temperature for the Voronoi model. Figure \ref{fig:angellLogLog} shows this data for both Voronoi \textit{and} vertex models in a log-log format. This plot suggests that at high temperatures the systems are much closer to having power-law scaling of the relaxation time with temperature than exponential scaling. Additionally, the dynamics themselves are fairly similar, with both models displaying stronger upturns for smaller values of the preferred shape parameter. We comment that there is not a strong signature of the zero-temperature $q_c\approx 3.81$ transition point even in the vertex model dynamical data here.

\section{Structure}

\begin{figure}
\includegraphics[width=0.85\linewidth]{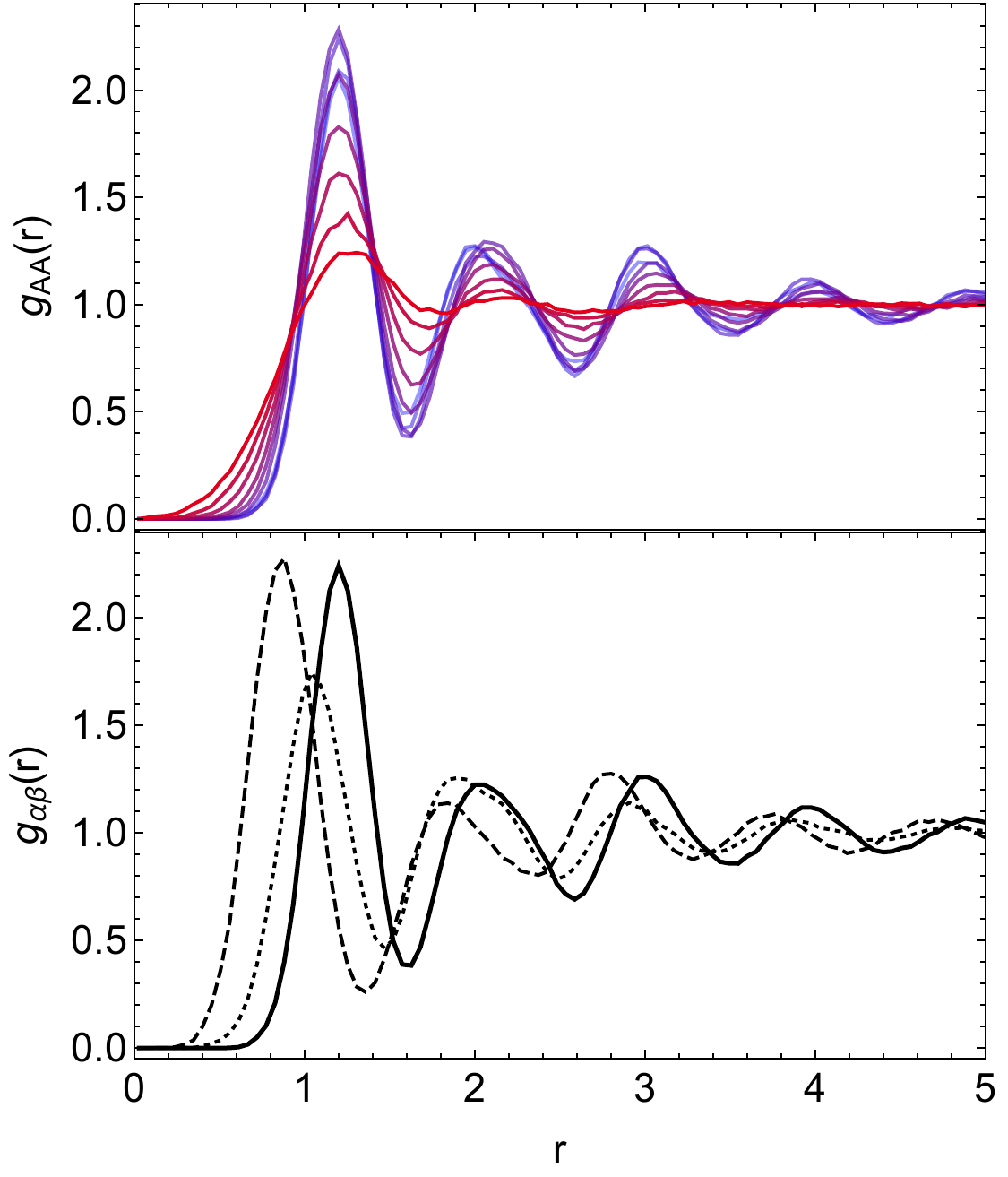}
\caption{\label{fig:gofr} (top) Radial distribution function for large cell-large cell correlations in systems of $N=1024$, $q_0=3.75$, and $T=1.05\times 10^{-3} - 1.5\times 10^{-1}$  (light blue to dark red). (bottom) Radial distribution functions for large cell-large cell correlations (solid), small cell-small cell correlations (dashed), and large cell-small cell correlations (dotted) for $N=1024$, $q_0=3.75$, and $T=1.05\times 10^{-3}$}
\end{figure}

For completeness, Fig. \ref{fig:gofr} shows examples of the radial distribution function of the Voronoi model at fixed $q_0$ as the temperature is varied. At low temperatures the particles become more strongly ordered, as might be expected for any liquid approaching a glassy phase. The degree of ordering (as reflected by, e.g., the change in the height of the first peak) is extremely modest as the dynamics slows down by orders of magnitude in the temperature regime probed in this figure. We comment that it would be interesting to try to connect the structural changes in this model with the dynamical changes reported above, for instance via a mode-coupling approach. The bottom frame of Fig. \ref{fig:gofr} also shows the different-species radial distribution functions for a particular state point at low temperature. This data, typical for what we have simulated, emphasizes that the model has stayed disordered, with no signature of local crystallization observed. 

\begin{figure}
\includegraphics[width=0.85\linewidth]{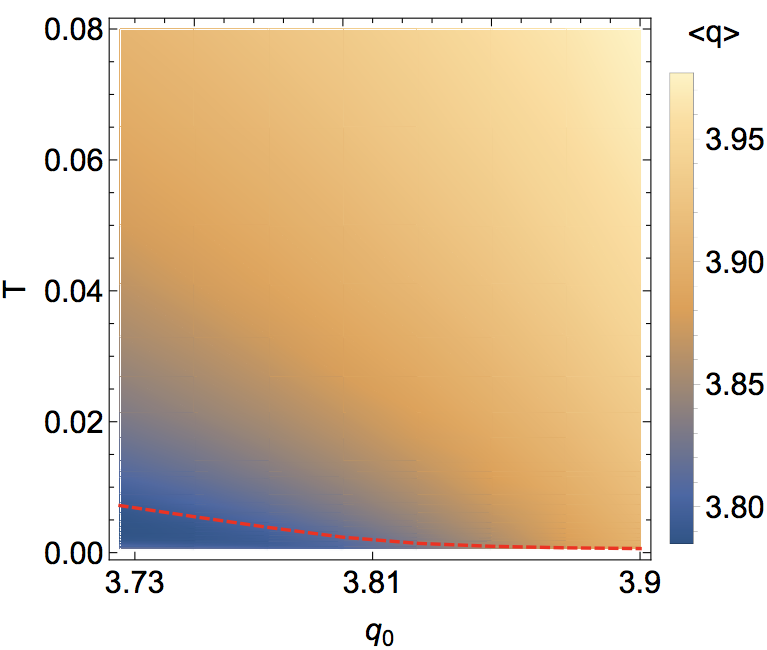}
\caption{\label{fig:linearPhasePlot} Comparison of the dynamical phase boundary and the structural measure of the typical shape of the cells. The color map corresponds to the mean measured shape parameter, $\langle q\rangle$, as a function of $q_0$ and $T$. The dashed red line corresponds to a line of constant $\tau_\alpha=10^4$.}
\end{figure}

As a complement to the phase diagram presented in the main text, we also plot the phase diagram showing both structural and dynamical measures on a linear scale. Shown in Fig. \ref{fig:linearPhasePlot}, this helps contextualize the low-temperature regime in which a strong discrepancy between a structural and a dynamical determination of the phase boundary can be seen. For comparison with the non-equilibrium case, see the plots in Ref. \cite{Bi2016}.

\section{Monodisperse Voronoi model}
\begin{figure}
\includegraphics[width=0.85\linewidth]{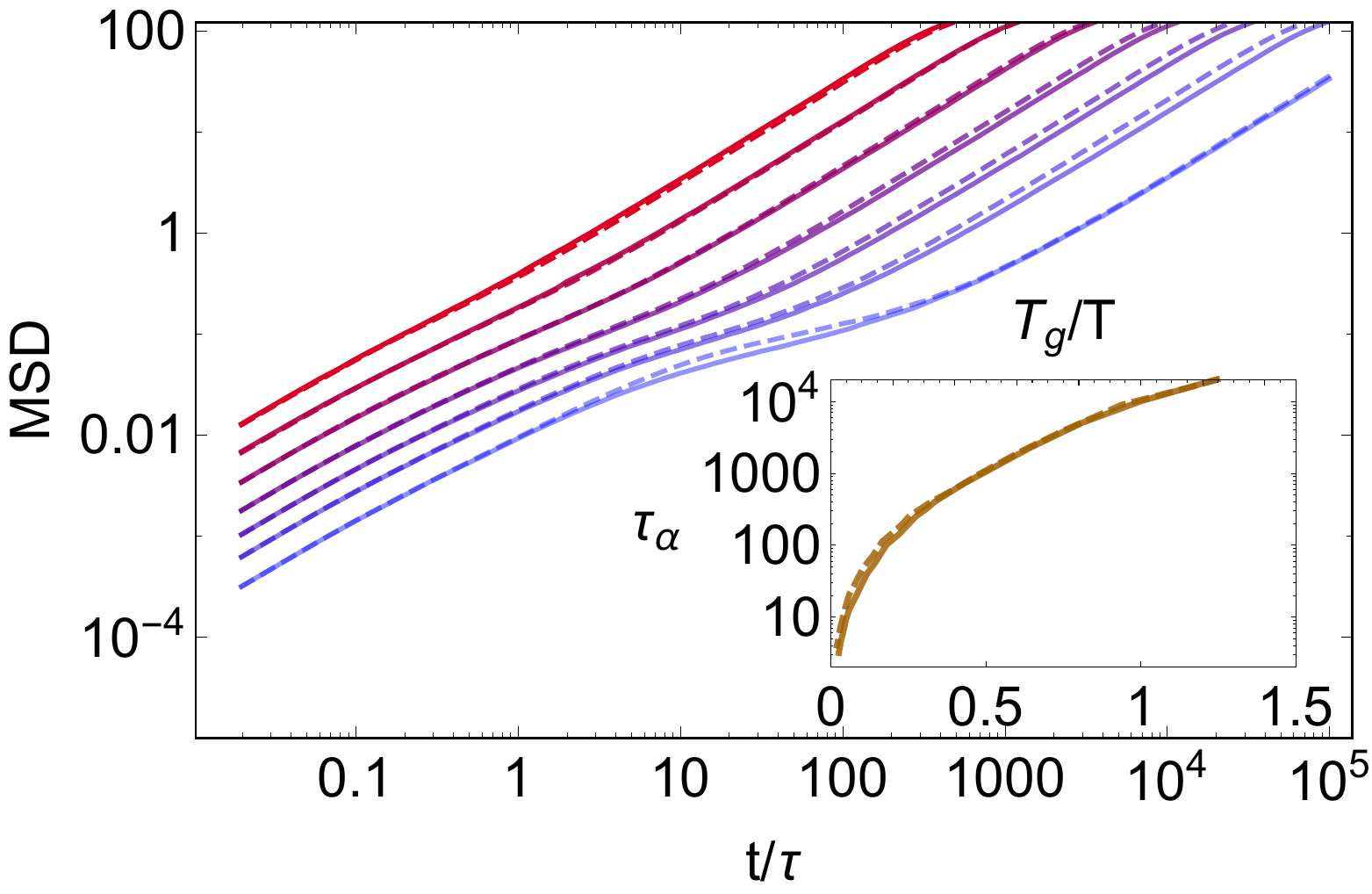}
\caption{\label{fig:monoBidisperseComp} Mean squared displacement vs time for bidisperse (solid curves) and monodisperse (dashed curves) Voronoi models at $q_0 = 3.8$ and $T=3.85\times 10^{-3} - 1.5\times 10^{-1}$  (light blue to dark red). [inset] Angell plot representation of $\tau_\alpha$ for the bidisperse (solid curve) and monodisperse (dashed curve) Voronoi models at $q_0 = 3.8$. The relaxation times are extremely similar, and notably the sub-Arrhenius behavior is still strikingly present.}
\end{figure}

In the main text we emphasized that we have performed simulations of bidisperse mixtures of Voronoi cells with different preferred areas (at constant preferred shape parameter, $q_0$). In order to facilitate comparisons with existing literature data, which have focused on monodisperse mixtures, we briefly present dynamical data comparing the bidisperse and monodisperse mixtures in a particular parameter regime where the monodisperse systems do not have a strong tendency to crystallize. Figure \ref{fig:monoBidisperseComp} shows both mean-squared displacement data and structural relaxation times for bidisperse and monodisperse systems at $q_0 = 3.8$ over a temperature range where the relaxation time is $\tau_\alpha \lesssim 10^4$. There are certainly small differences (indeed, there is no reason the dynamics should be identical), but the strong similarities observed suggest that our general results are very robust to the particular choice of bidispersity we made.

\end{document}